# Counterfactual computation revisited

Onur Hosten, Matthew T. Rakher, Julio T. Barreiro, Nicholas A. Peters, and Paul Kwiat

*Department of Physics, University of Illinois at Urbana Champaign, Urbana, Illinois 61801-3080*
(Dated: June 26, 2006)
hosten@uiuc.edu

Mitchison and Jozsa recently suggested [1] that the 'chained-Zeno' counterfactual computation protocol recently proposed by Hosten et al. [2] is counterfactual for only one output of the computer. This claim was based on the existing abstract algebraic definition of counterfactual computation [3], and indeed according to this definition, their argument is correct. However, a more general definition (physically adequate) for counterfactual computation is implicitly assumed in [2]. Here we explain in detail why the protocol *is* counterfactual and how the 'history tracking' method described in [1] inadequately represents the physics underlying the protocol. Consequently, we propose a modified definition of counterfactual computation. Finally, we comment on one of the most interesting aspects of the error-correcting protocol [2].

By definition, counterfactual means 'contrary to fact', and counterfactual computation refers to a process in which one obtains information about the solution of a computational problem contrary to the fact that the computer producing the answer did not run in the process. This is possible only in the realm of quantum computers, where the computer can exist in a quantum superposition of 'running' and 'not running' at the same time. As a physical example, computer 'running' can mean that a photon is actually passing through a box of circuits which imprints the answer to a computational problem on some properties of the photon. Computer 'not running' in this case would mean that the photon is not passing through this physical box, instead following an entirely different path. In this case, a counterfactual computation would be the process of obtaining information about the answer of the computational problem, even though the photon did not pass through the physical box of circuits. There is no debate about any of the statements up to this point.

## The controversy

The controversy arises over the particular definition of what constitutes a counterfactual outcome. In their original paper [3] (and reiterated in [1]), Mitchison and Jozsa define a counterfactual process to be one in which there are *no* possible histories or amplitudes in which the process could have occurred. We admit that, *under such a definition*, which initially seems very reasonable, the chained-Zeno protocol is not counterfactual for some of the computer outputs. However, as we shall describe here, we believe that this original definition is too restrictive (or from a different perspective, not restrictive enough!) to properly accord with the physical situations arising in our scheme for counterfactual computation.

Before going into detailed descriptions, we would like to summarize the essence of the controversy, in our words, with the following gedanken experiment (Fig. 1). In the experiment, the amplitude of a photon is split into two via a 50/50 beamsplitter (BS); part of the amplitude goes directly to a screen, and part of it goes to an interferometer. The interferometer is aligned so that there is complete destructive interference in the bottom port, and this dark port also leads to the screen (see the caption of Fig. 1). The question is whether or not a photon detected at the screen passes through the point labeled 'C' in the interferometer before it arrives at the screen. Our answer is that it does *not*, because no photon exits the dark output port of an interferometer, due to the destructive interference of the two possible trajectories of the photon.

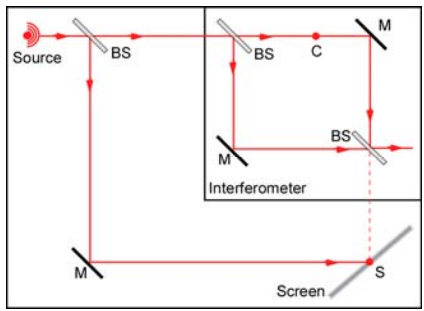

**Figure 1** The gedanken experiment. M: Mirror. The amplitudes in the two arms of the interferometer (i.e., the upper arm containing the point 'C', and the lower arm which does not contain the point 'C') interfere destructively at the interferometer output port which leads to the screen (this path is shown in dotted lines). Therefore, no photon ever comes out of this output port. If we detect a photon at point 'S' on the screen, then this photon definitely did not come from the interferometer, and therefore did not pass through the point 'C'.

The controversy comes about when we attempt to explain the physics with individual histories (as the existing abstract definition of counterfactual computation is based on this). We encounter three distinct histories (Fig. 2), or quantum trajectories, in the outcome in which the photon is detected at point 'S' [4].

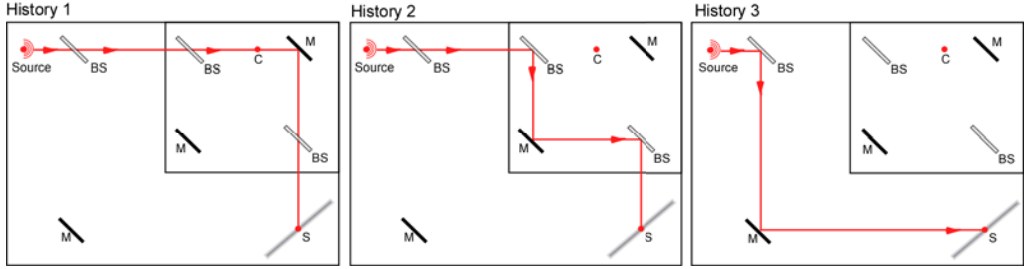

**Figure 2** Individual quantum histories contributing to the detection of the photon at point S.

'History 1' involves an event in which the photon passes through point 'C'. To make the full analogy: the photon passing through the point 'C' corresponds to the computer running. We interpret Mitchison and Jozsa's argument [1], applied to this physical setup, to say that the photon *does* pass through point 'C' before it arrives at the screen (i.e., the computer runs – the computation is not counterfactual – in the outcome in which the photon is detected at 'S'), because there is a history in which this occurs, i.e., 'History 1' in Fig. 2. [5] However, any interpretation about the physics behind the overall event (e.g., whether the photon passes through point 'C' before it is detected at the screen) should *not* be made using individual histories (or a partial group of histories), but instead only after coherently superposing all possible solutions, i.e., histories. The fact that the interferometer output port that leads to the screen is dark arises from the destructive interference between 'History 1' and 'History 2': The quantum probability amplitudes corresponding to 'History 1' and 'History 2' are equal in magnitude and opposite in sign, so they exactly cancel each other. The conclusion is that only 'History 3' contributes to the outcome in which the photon is detected at point 'S', and in this outcome there is no relevant history with the photon traveling through point 'C'. [8]

**Single-photon implementation of the chained-Zeno protocol**

Now we proceed to address the arguments in the context of counterfactual computation in more detail. We will use a simpler-to-digest version of the optical implementation of the 'chained-Zeno' version in [2] in order to discuss the problem on physical grounds. Consider a computer with two possible answers, '0' and '1'. Also, consider that the computer has an *on/off* switch, and it can run only if the switch is *on*. Moreover, consider that the computer switch itself is connected to another *on/off* switch (we will call this the 'subroutine switch' for reasons to become apparent), meaning that the computer can run only if both of the switches are *on*. This system can be described by the states of three qubits $|i\rangle_1|j\rangle_2|k\rangle_3$, where indices are the qubit numbers, and $i, j, k = 0$ or $1$. The first qubit represents the 'subroutine switch', the second qubit represents the 'computer switch', and the third qubit represents the 'computer input/output register'. The computer outputs the answer to the third qubit as $|0\rangle_3$ or $|1\rangle_3$, only if the state of the first two qubits is $|1\rangle_1|1\rangle_2$. (If either qubit 1 or 2 is in state $|0\rangle$, the state of the third qubit is not affected by the computer.)

The single-photon optical implementation of the computer (comp) is shown below. If both the 'subroutine switch' and the 'computer switch' are *on*, then the photon enters the computer; if the yet-unknown answer to the computation is '0', the photon is transmitted (Fig 3.a), and if the yet-unknown answer is '1', the photon is reflected downwards (Fig 3.b). However, if either one of the switches is *off*, then the photon follows an entirely different path and does *not* enter the computer, as shown symbolically in Fig. 3.c. (Readers interested in this single-photon multi-qubit type encoding can find explanatory examples in [6].)

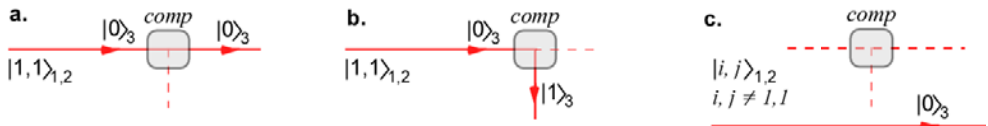

**Figure 3** Single-photon optical implementation of the computer

Since [1] argues that the final result is *not* counterfactual only when the computer output is '0', we shall concentrate only on this case. (We all agree that the computation is counterfactual in [2] when the computer output is '1'.) In what follows, we will investigate the first cycle of the 'routine' of the procedure. The initial state of the system is $|0\rangle_1|0\rangle_2|0\rangle_3 \equiv |000\rangle$. The 'routine' starts with the rotation $R'$ on the 'subroutine switch' [7]. Conditional on the 'subroutine switch' being $|1\rangle_1$, a subroutine runs and applies $N$ cycles ($N=2$ for the example we describe here) of the rotation $R$ on the 'computer switch' [7] followed by running the computer only if the 'computer switch' is $|1\rangle_2$. After each insertion of the computer, the value of the computer output (qubit 3) has to be measured to see if the computer has run in case the answer to the computation is $|1\rangle_3$. Similarly at the end of the subroutine, qubit 2 has to be measured eventually to figure out whether the subroutine has run at all. The single-photon optical implementation is given below. Different paths, as labeled, serve as different states, and beamsplitters (BS) with the correct reflection and transmission coefficients serve as the rotation operators by mixing the input modes.

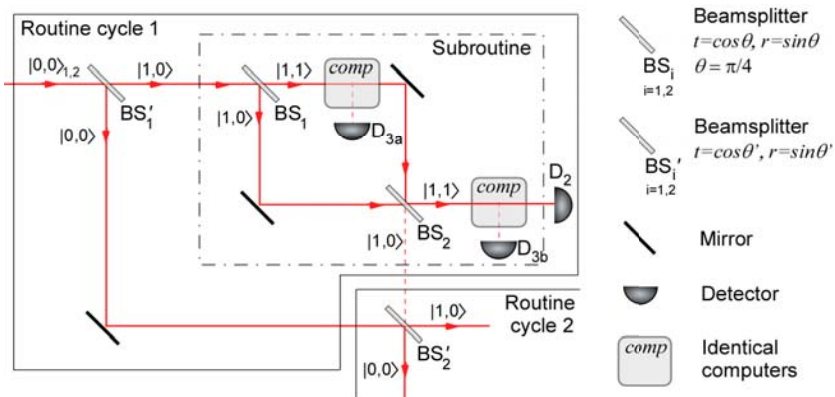

**Figure 4** Single-photon optical implementation of the first cycle of the 'routine'. Primed beamsplitters BS′ implement the rotation $R'$ on the first qubit, and the unprimed beamsplitters BS implement the rotation $R$ on the second qubit.

Measurements of computer output, i.e., qubit 3, are performed by detectors $D_{3a}$ and $D_{3b}$. Detection of a photon at these detectors indicates that the photon has gone through the computer, i.e., the computer has run, and the outcome of the computation is '1'. If no detection occurs at $D_{3a}$ or $D_{3b}$, we do not know whether or not the photon has passed through the computer, but we can say that the state of qubit 3 is not $|1\rangle_3$, i.e., qubit 3 is in state $|0\rangle_3$. Similarly, the measurement of qubit 2 at the end of the subroutine is performed by detector $D_2$. Detection of a photon at this detector indicates that the photon has gone through the subroutine, i.e., the subroutine has run.

Referring to Fig. 4, the photon incoming to the first cycle of the 'routine' is initially put into a superposition of traveling paths $|0\rangle_1$ and $|1\rangle_1$ via BS'$_1$. Path $|1\rangle_1$ contains the subroutine, while path $|0\rangle_1$ does not contain anything. The amplitude that enters the subroutine is further split into a 50/50 superposition of going through paths $|0\rangle_2$ and $|1\rangle_2$ via BS$_1$. Path $|1\rangle_2$ contains the computer. Since we are considering only the case where the answer to the computation is '0', the computer transmits the entering amplitude (see Fig. 3a), and there is no chance of detector $D_{3a}$ clicking. Paths $|0\rangle_2$ and $|1\rangle_2$ are then combined on BS$_2$ where the amplitudes interfere to give complete constructive interference at the new $|1\rangle_2$ path, and to give complete destructive interference at the new $|0\rangle_2$ path. This latter path is shown in dashed lines in the optical implementation to stress that there is no probability for the photon to be found on this path, consequently, there is no probability *flux* on this path. There are two possible ways in total to enter the second cycle of the routine; the dashed path coming from the subroutine (labeled $|1,0\rangle$) and the path labeled $|0,0\rangle$. In the outcome in which the photon makes it to the second cycle of the 'routine' (which means that detector $D_2$ does not click) we know that it cannot have come from the subroutine since there is no probability flux or probability amplitude in this dashed path. One could in principle verify this by putting a detector at the end of the dashed path before the second cycle of the 'routine'; if the computation result is '0', this detector would *never* register a photon. Therefore, we conclude that in the outcome in which the photon makes it to the second cycle of the 'routine', the photon is definitely coming from the path labeled $|0,0\rangle$, and as a consequence it has not passed through the computer. That is, the first cycle of the 'routine' is completely counterfactual. Since all the 'routine' cycles are identical, the *entire* protocol is counterfactual.

The abstract evolution of the first 'routine' cycle is given below for reference. Note that the computer will run only if the state is $|110\rangle$. For convenience, we have underlined the terms which have some amplitude passing through the computer in their histories.

$$|000\rangle \xrightarrow{R'} \cos\theta'|000\rangle + \sin\theta'|100\rangle$$

$$\xrightarrow{R} \cos\theta'|000\rangle + \frac{\sin\theta'}{\sqrt{2}}\left(|100\rangle + |110\rangle\right) \xrightarrow{comp} \cos\theta'|000\rangle + \frac{\sin\theta'}{\sqrt{2}}\left(|100\rangle + \underline{|110\rangle}\right) \quad , \text{measure } 3 \rightarrow 0_3$$

$$\xrightarrow{R} \cos\theta'|000\rangle + \sin\theta'\underline{|110\rangle} \xrightarrow{comp} \cos\theta'|000\rangle + \sin\theta'\underline{|110\rangle} \quad , \text{measure } 3 \rightarrow 0_3$$

measure $2 \rightarrow 0_2$ or $1_2$ .

In the abstract evolution above, the measurement on qubit 3 in the second line (third line) corresponds to the measurement made by detector $D_{3a}$ ($D_{3b}$) in the optical implementation: Assuming these detectors do *not* click, we know that qubit 3 is in state $|0\rangle_3$. Similarly, the measurement on qubit 2 in the last line corresponds to the measurement made by detector $D_2$. Note that the measurement outcome '$0_2$' on the last line ($D_2$ did *not* fire) would indicate that the photon has successfully made it to the second cycle of the 'routine' (here we are not interested in the measurement outcome '$1_2$').

### Individual histories and the abstract definition of counterfactual computation

The available abstract definition of counterfactual computation [1], however, seems to indicate that this protocol is not counterfactual. To understand the discrepancy, we have to refer to the individual histories (or quantum paths) leading to the second cycle of the 'routine'. There are three distinct histories to end up in the second cycle, as shown in Fig. 5.

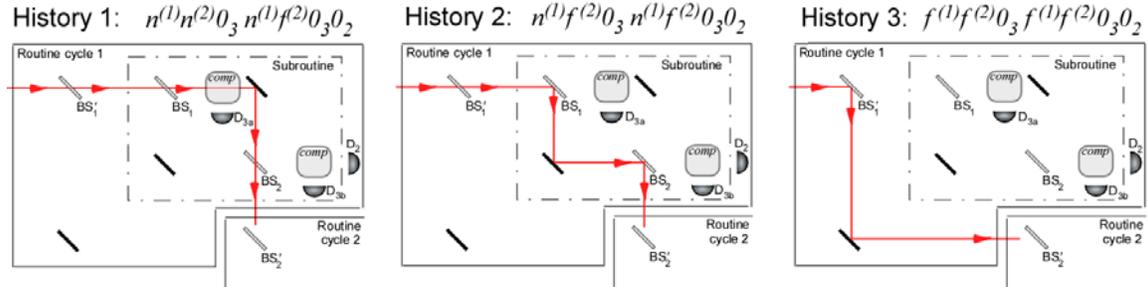

**Figure 5** Individual histories contributing to the outcome in which the photon ends up in the second cycle of 'routine'.

We use a notation similar to the one introduced by Mitchison and Jozsa [1]. The labels of the histories are indicated above them. $n^{(1)}$ ($f^{(1)}$) indicates whether the 'subroutine switch' is *on* (*off*), i.e., whether the photon is *passing* (*not passing*) through the subroutine. Similarly $n^{(2)}$ ($f^{(2)}$) indicates whether the 'computer switch' is *on* (*off*), i.e., whether the photon is *passing* (*not passing*) through the computer. All combinations are possible except $f^{(1)}n^{(2)}$ since the computer *cannot* run if the subroutine is *not* running. $0_3$ or $1_3$ in the history labels would indicate the outcome of a measurement made on qubit 3, i.e., the outcomes of the measurements made by detectors $D_{3a}$ or $D_{3b}$. Similarly $0_2$ or $1_2$ would indicate the outcome of a measurement made on qubit 2, i.e., the outcome of the measurement made by detector $D_2$. To see the reasoning behind the labels, consider 'History 1' in Fig. 5. In this history, the photon is transmitted by $BS'_1$, so that the subroutine is *on*; the photon subsequently is transmitted by $BS_1$, so the computer is also *on*. Thus, symbolically the history starts with $n^{(1)}n^{(2)}$. Next, the photon goes through the computer, and since detector $D_{3a}$ doesn't register the photon, which would have meant $1_3$, we infer that the measurement on qubit 3 is '0'. Therefore, the history continues with $0_3$. After this, the photon is transmitted by $BS_2$, so this time the computer of *off*, although the subroutine is still *on*. This brings a $n^{(1)}f^{(2)}$ symbol to the history. Now, the photon doesn't go through the computer, and again, measurement of qubit 3 (i.e., the 'computer input/output register') gives '0' since this time detector $D_{3b}$ doesn't register the photon. This is a $0_3$ again for the history label. Finally, since the photon goes straight to 'Routine cycle 2', i.e., towards $BS'_2$, the photon is not registered at detector $D_2$, from which we infer that the measurement on qubit 2 is '0'. Therefore, we add a $0_2$ label to the history. Putting all the parts together we get 'History 1': $n^{(1)}n^{(1)}0_3n^{(1)}f^{(2)}0_30_2$. Similar arguments apply for Histories 2 and 3.

In the abstract notation we generate a table (Table-I) of three histories (similar to Table-I in [1]) which yield the same real measurement outcomes $0_30_30_2$. This measurement outcome set indicates that the photon makes it to the second cycle of the 'routine'. In Table-I, $h$ refers to a particular history and $v_h$ refers to the quantum state vector corresponding to that history. There is one-to-one correspondence between Fig. 5 and Table-I.

**TABLE-I**

|    | $h$ | $v_h$ |
|----|-----|-------|
| 1: | $n^{(1)}n^{(2)}0_3 n^{(1)}f^{(2)} 0_30_2$ | $-\sin\theta'\left|100\right\rangle/2$ |
| 2: | $n^{(1)}f^{(2)}0_3 n^{(1)}f^{(2)} 0_30_2$ | $\sin\theta'\left|100\right\rangle/2$ |
| 3: | $f^{(1)}f^{(2)}0_3 f^{(1)}f^{(2)} 0_30_2$ | $\cos\theta'\left|000\right\rangle$ |

The fact that there is no net probability amplitude entering 'Routine cycle 2' from the path exiting the 'subroutine' arises from the destructive interference between 'History 1' and 'History 2': The quantum probability amplitudes corresponding to 'History 1' and 'History 2' are equal in magnitude and opposite in sign, and they exactly cancel each other. The conclusion is that only 'History 3' contributes to the outcome in which the photon ends up in 'Routine cycle 2', and in this outcome there is no history with the photon passing through the subroutine, and as a result, no chance of passing through the computer.

Earlier, we stressed that basing any interpretation concerning physical quantities or observations on individual histories or a partial group of histories can be misleading. This is the disadvantage of the existing abstract description of counterfactual computation when applied to the 'chained-Zeno' protocol: It does not have enough variables to physically explain the protocol, and as a consequence makes an interpretation based on only a partial group of histories. Specifically, there is then only one variable for labeling paths, that is $n/f$ (short for *on/off*), which indicates whether the amplitude traveled through the computer or not. This method yields a history table (Table I in [1]) whose relevant lines are given below in Table-II.

**TABLE-II**

|    | $h$ | $v_h$ |
|----|-----|-------|
| A: | $n\, 0_3 f\, 0_30_2$ | $-\sin\theta'\left|100\right\rangle/2$ |
| B: | $f\, 0_3 f\, 0_30_2$ | $\cos\theta'\left|000\right\rangle + \sin\theta'\left|100\right\rangle/2$ |

We see that histories '2' and '3' of Table-I are grouped together to give the second line of Table-II. The corresponding pictorial representations are given in Fig. 6.

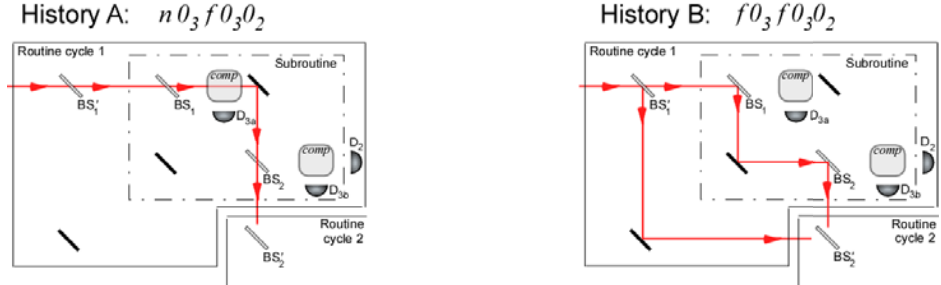

**Figure 6** 'History 2' and 'History 3' of Fig. 5 are grouped together to give a new 'History B', in the approach of [1].

As is evident from Fig. 6, when the two histories 'A' and 'B' are treated and interpreted separately, one reaches a misconception. It looks as if 'History B' is the counterfactual part of the protocol that ends up in the second cycle of the 'routine', but there is also a contribution from 'History A', in which the photon travels through both the subroutine and the computer. However, we know that nothing comes from the subroutine due to the destructive interference of the two subroutine-running trajectories in 'History A' and 'History B' combined. To repeat: A detector placed in the upper entrance port of BS$'_2$ would *never* detect a photon. Therefore, we feel justified in claiming that the photon did not pass through the subroutine (since we also are considering only cases where none of the detectors $D_{3a}$, $D_{3b}$ or $D_2$ detects the photon).

Here we repeat the existing definition given in [1] for a counterfactual computation:
A set *m* of measurement outcomes is a counterfactual outcome if (1) there is only one history associated to *m* and that history contains only *f*'s, and (2) there is only a single possible computer output associated to *m*.

In light of the above discussion, we propose to redefine a counterfactual computation as follows [8]:
(1) Identify and label all histories (quantum paths), with as many labels as needed, which lead to the same set *m* of measurement outcomes, and (2) coherently superpose all possible histories. (3) After canceling the terms (if any) whose complex amplitudes together add to zero, the set *m* of measurement outcomes is a counterfactual outcome if (4) there are no terms left with the computer-running label in their history labels, and (5) there is only a single possible computer output associated to *m*.

In our example, the computer-running label relevant to item (4) is $n^{(2)}$.

## On the error-correcting protocol

Finally, we would like to comment on the conclusions of [1] concerning the error-correcting protocol in [2]. Reference [1] compares the difference in the information gain from the computer when it is simply run $2NN'$ times, and when the error-correcting chained-Zeno protocol is run with the parameters $N$ and $N'$ – since $2NN'$ insertions of the computer are required in either case. This was something that was not calculated in [2] and is definitely informative. In some circumstances we agree with Mitchison and Jozsa that the former technique has advantages (especially since it is much simpler to implement!). However, one of the potentially important aspects of any quantum computing protocol involves sending the output to *another* quantum computer. Decoherence in the computer causes it to output a partially mixed (or decohered) state; hence, if one intended to feed the answer coming from the algorithm *coherently* to another quantum computation, there would be a propagation of errors. In contrast, the error-correcting protocol described in [2] outputs an extremely pure answer, which may be sent coherently into another quantum computation, thereby avoiding the propagation of errors. This is one potential advantage to counterfactually performing small parts of a larger computation.

## On the 'Random guessing limit'

Before concluding, we would like to make a clarification, based on footnote 5 of [1]. Specifically, we wish to clear up a verbal confusion on the argument about the 'random guessing limit', which we confess not to be very clear in [2]. The random guessing limit actually refers to $p_0+p_1 \leq 1$ in [1]. If the answer is 1, the 'Zeno' scheme (not the 'chained-Zeno') reveals that the answer is 'not 0' without the computer running.

But if the answer is 0, the computer runs. Because one does not know ahead of the time which is the answer, the probability of the computer running is ½. The goal is to *not* run the computer yet still obtain the right answer. Using the Zeno scheme one achieves this goal ½ of the time; the same as random guessing (don't run the computer, just guess '0' or '1'). It is only in this sense that we intended to claim in [2] that the simple 'Zeno' scheme is no better than random guessing. Only by employing the chained-Zeno methods described in [2] can this limit be exceeded.

### Conclusions

We have shown why the 'chained-Zeno' counterfactual computation protocol of [2] is indeed counterfactual for the computer output '0', in response to the objections in [1]. We have identified why the original abstract description of counterfactual computation is at odds with the results, and we proposed a new definition that is more general and can properly accommodate the current protocol. Like Mitchison and Jozsa, we agree that the limits and the usefulness of the error-correcting protocol are not very clear. It is our hope that this discussion will stimulate further discourse on these and related topics.

# Reply to the response of Dr. Mitchison and Dr. Jozsa

In this reply, we address the double-slit analogy given by Mitchison and Jozsa [1], and resolve their worries about this analogy. Next, we show that their addition of a fourth register to prove that our protocol is not counterfactual is not legitimate. Then, we show that our own proposed definition making use of histories can sometimes give contradictory answers as to whether or not the computation is counterfactual; consequently, we must conclude that it is not an optimal definition. Nevertheless, we still argue that the chained-Zeno counterfactual computation (CFC) protocol is completely counterfactual. In the appendix, we show that under the Mitchison-Jozsa definition for CFC, even their original simple CFC protocol can be shown to give contradictory answers as to whether or not it is counterfactual, depending on the details of the analysis, rendering it not an optimum definition as well. We conclude with a brief discussion on the basis dependence of counterfactuality, and the search for a more descriptive definition of a counterfactual process.

### The double slit experiment analogy

In their response to us, Mitchison and Jozsa [1] give a double-slit analogy for the interferometer example of our Fig.1 and Fig. 2, adding: "we can regard the two slits as being analogous to the two arms of the interferometer in their Figure 2, and it is only by arguing that the photon goes through neither arm that they can claim that computer C does not run." Indeed, while we agree with this statement, our example in Figs.1&2 is more complicated than a double-slit alone. We have no contradiction with the usual interpretation of the standard double-slit experiment: One cannot claim which slit the photon went through only from a detection event on the screen. Also, claiming that 'the photon took neither of the slits to reach a dark fringe on the screen' would be confusing, since there is no such outcome to begin with, i.e., one never registers a photon at a dark fringe. The clarification of the full analogy in terms of slits is depicted in Fig. 7a.

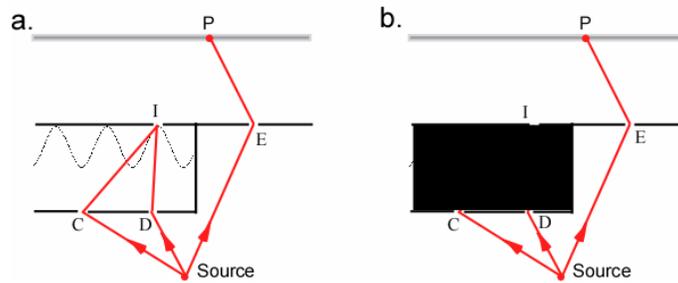

**Figure 7** The full analogy in terms of slits, as opposed to interferometers as in Figs.1&2.

Slits C and D form the double-slit part of the analogy. The photon passing through slit C corresponds to the computer running. The interference pattern of the double slit is depicted as a wave in Fig 7a. We place another slit (slit I) at one of the dark fringes. Since a photon will never end up at a dark fringe, no photon will come out of slit I to go to point P on the screen [9]. This can be verified by placing a detector in front of slit I (this corresponds to adding the fourth register which is incremented every time the *subroutine* runs, as we discuss in the next section). We can also draw the experiment as in Fig. 7b. What is inside the black box is irrelevant. What is important is that no photon ever comes out of the box (a photon entering the box gets absorbed at some place on the walls of the box), and that the computer C is inside the box. Now the only way that a photon can be detected at point P on the screen is if it traveled through slit E. Another way of saying this is that the photon goes through neither slit C nor D when it is detected at point P. Therefore, in this outcome the computer does not run.

If instead we put a which-path detector (even one that does not destroy the photon in the process) before slit C in Fig. 7a (this corresponds to adding the fourth register which is incremented every time the *computer* runs, as Mitchison and Jozsa proposed; we address this in more detail below), there will no longer be interference between the slits. In the absence of a dark fringe at slit I, we now cannot claim that a photon detected at point P did not travel via slit C. Thus, in order for the protocol to work properly, we must avoid such a which path detector. In the context of counterfactual computation this means that the state of the computer switch qubit should not be entangled with any ancillary or workspace qubits of the computer for CFC to work properly.

### On the addition of a fourth register

Mitchison and Jozsa [1] highlight their concerns with our new definition for counterfactual computation by adding a fourth register initially set to $|0\rangle_4$, and flipped to $|1\rangle_4$ (or "incremented", as Mitchison and Jozsa phrase it) whenever the computer runs. Specifically they claim that if the computation were "counterfactual without the fourth register, it should remain so with this register added". In fact this seemingly innocuous statement is false. Their modification is equivalent to putting a wave-plate (after point 'C' in Fig. 1) which rotates the polarization of the photon orthogonal to its initial polarization, so that the amplitude on this path cannot show destructive interference with other paths when combined on BSs. In other words, this adds distinguishing information between the 'running' and 'not-running' histories of the computer. However, *the indistinguishability of these two histories, when the answer to the computation is $|0\rangle_3$, is what lies at the heart of counterfactual computation,* by giving rise to the destructive interference at the bottom output port of the interferometer in Fig. 1. It is worth noting that with such a fourth register, even their original simple version of counterfactual computation [3, 2] does not work. If we want to constrain ourselves to protocols that actually work, such modifications should be avoided. However, as we explain in the next paragraph, we would like to use a similar modification, specifically to add the fourth register in a different way, to show that the computation can be shown to be "counterfactual".

Consider adding to our protocol a fourth register, initially set to $|0\rangle_4$, and incremented every time the *subroutine* runs (once at the beginning of the subroutine cycles). [Note the difference with Mitchison and Jozsa's extra register, which was incremented directly by the computer.] This also corresponds to flipping the polarization of the photon before it enters the interferometer in Fig. 1. When the computer output is $|0\rangle_3$, in place of our Table-I in this article and their Table-IV in [1], we have Table-III. As can be seen from Table-III, not only do History 1 and History 2 cancel out, but we could also tell from a further measurement on the fourth register at the very end, whether or not the subroutine has run. In the set of measurement outcomes $0_3 0_3 0_2$, the measurement on the fourth register at the very end can only give $0_4$, indicating that the subroutine has not run. Since the computer running is conditioned on the subroutine running, the computer *does not* run either. However, we repeat that the addition of any such fourth register should be avoided since then CFC does not work, i.e., the computer-not-running condition is satisfied, but we cannot infer the answer.

**TABLE-III**

|  | $H$ | $v_h$ |
|---|---|---|
| 1: | $n^{(1)}n^{(2)}0_3\,n^{(1)}f^{(2)}\,0_3 0_2$ | $-\sin\theta'\,|1001\rangle/2$ |

| 2: | $n^{(1)}f^{(2)}0_3 n^{(1)}f^{(2)}0_3 0_2$ | $\sin\theta' |1001\rangle/2$ |
|---|---|---|
| 3: | $f^{(1)}f^{(2)}0_3 f^{(1)}f^{(2)}0_3 0_2$ | $\cos\theta' |0000\rangle$ |

## Problems with definitions making use of histories

In fact, as was brought to our attention by Jerry Finkelstein [10], our own new definition for CFC (introduced on page 6 of this article) does not give a unique answer in all cases as to whether or not the computation is counterfactual, making it a less-than-optimal definition. Not very surprisingly, the Mitchison-Jozsa definition of CFC can suffer the same problem, even in explaining their original simple CFC scheme, depending on the analysis one uses to examine the scheme. In this section we concentrate on our new definition only, and in the appendix we discuss the Mitchison-Josza definition. Let us illustrate the problem by modifying Fig. 1. We replace the screen in Fig. 1 with another BS ($BS_2$) which acts as a which-path-information eraser (see Fig. 8). This means that in the outcome in which the photon is detected at point S in Fig. 8, one could not tell whether the photon was coming from the interferometer (which is boxed and labeled as 'interferometer'), or from the lower path that also leads to $BS_2$. We ask the question: Did the photon pass through point C prior to its detection at point S?

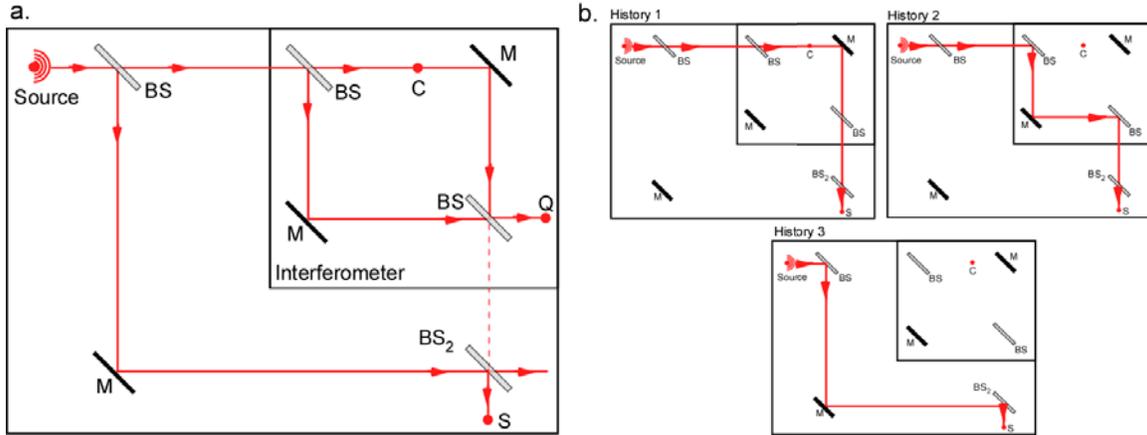

**Figure 8 a.** Modified version of the gedanken experiment in Fig. 1. **b.** Individual quantum histories contributing to the detection of the photon at point S.

Looking at Fig.8a we claim that if the photon enters the interferometer, it has to be detected at point Q, since no amplitude leaves the interferometer from its bottom output port due to the destructive interference. Consequently, *based on this physics argument,* if the photon is detected at point S, it had to follow the path shown in History 3 of Fig 8b, since no light can come from the bottom output port of the interferometer due to the destructive interference.

However, consider what happens if we apply the definition that we proposed earlier for a counterfactual process, in which we are to add together amplitudes from different histories first. It is important that there is no distinguishing information about the photon at point S in any of the three histories, so that we can safely add the amplitudes. Call the amplitudes corresponding to History 1, History 2 and History 3, $c_1$, $c_2$ and $c_3$ respectively. We can choose the values of these numbers by adjusting individual BS reflectivities and phase conventions on them. One constraint is that $c_1 = -c_2$, due to the destructive interference at the bottom output port of the interferometer. Consider the case: $c_1 = a$, $c_2 = -a$, $c_3 = a$. Before we argued that History 1 and History 2 destructively cancel each other, and the only contribution to detection at point S comes from History 3, i.e., the photon *does not* pass through point C prior to its detection at point S. On the other hand, with our definition it might seem that one could equivalently say that History 2 and History 3 destructively cancel each other, and the only contribution to detection at point S comes from History 1, i.e., the photon *does* pass through point C prior to its detection at point S.

The two interpretations obviously contradict with each other, which means that individual histories connecting the beginning and the end of this experiment *cannot* be used to answer which path the photon took. It is like the question 'which one of the three slits did the photon take in a simple triple-slit experiment before it is was detected at the screen', a fundamentally meaningless question. However, in an experiment implementing Fig. 8 (or the chained-Zeno protocol), the amplitudes do not interfere "all at once". The experiment happens in steps, interfering amplitudes in pairs on BSs in real time. In this case, the analysis should be made in steps as well, and the question 'Did the photon pass through point C?' *is* meaningful. Because there is no chance of a photon following the dashed path in Fig. 8, we can legitimately infer that the photon did not pass through point C prior to its detection at point S, (i.e., the chained-Zeno CFC is then fully counterfactual).

### Conclusions 2

In summary, we conclude that in a physical implementation of the chained-Zeno CFC, one obtains the answer without the computer running (physical manifestation of the "computer not running" depends on the physical objects used in the experiment) for the computer output $|0\rangle_3$. This is justified by the interferometer analogy that we presented, because the photon *does not* pass through point C prior to its detection at point S, since it *does not* pass through the interferometer at all. In contrast, the original definition given by Mitchison and Jozsa [3, 1] incorrectly predicts that a photon *does* travel along the dashed lines on its way to point S in Fig. 8a (because this is the only way to go from point C to point S). We have found that definitions or interpretations that use histories connecting the beginning and the end of the protocol (including our own proposed definition in the first part of the article) may be inadequate, because they can lead to ambiguous answers.

In order to determine whether a process is counterfactual or not, one needs to make the analysis in steps, calculating the amplitudes that are output in different states of a quantum operation (or gate), which are in turn going to become the input states to other gates (or operations). Therefore, the choice of basis states becomes crucial in answering whether or not any amplitude is entering the gate: A counterfactual process viewed in one basis set may not be counterfactual in a different basis set. Consequently, an abstract, basis-independent definition of a counterfactual process may become impossible. However, in an experiment, the physical meaning of computer not running is usually clear, and the physical experiment fixes the basis set for the interpretation. We are still seeking a more descriptive definition of counterfactual processes.

### APPENDIX: Analysis of the original simple CFC protocol

Interestingly enough, according to the definition of Mitchison and Jozsa [3, 1], even their original simple CFC protocol can be shown to give contradictory answers as to whether or not it is counterfactual, depending on how one makes the analysis. Let us summarize this protocol. We have a computer with an input/output register qubit which can take on values $|0\rangle_2$ or $|1\rangle_2$ depending on the answer to the computation, and we have a computer switch with values $|0\rangle_1$ and $|1\rangle_1$, corresponding to the switch being 'Off' and 'On', respectively. The initial state is $|0\rangle_1|0\rangle_2$. One puts the computer in a superposition of running and not running by the unitary operation $U_{|0\rangle_2}$: $|0\rangle_1 \to (|0\rangle_1+|1\rangle_1)/\sqrt{2}$ and $|1\rangle_1 \to (-|0\rangle_1+|1\rangle_1)/\sqrt{2}$, on the switch. The subscript $|0\rangle_2$ next to $U$ indicates that the operation is conditional on the input/output register being $|0\rangle_2$. Following this, the computer is allowed to run only if the switch qubit is $|1\rangle_1$. As the final step of the protocol, one applies the same unitary operation $U_{|0\rangle_2}$ one more time. If the computer performs the operation $|0\rangle_2 \to |0\rangle_2$ (i.e., if the answer is '0'), then the evolution of the system is $|0\rangle_1|0\rangle_2 \to |1\rangle_1|0\rangle_2$. On the other hand, if the computer performs the operation $|0\rangle_2 \to |1\rangle_2$ (i.e., the answer is '1'), then the evolution of the system is $|0\rangle_1|0\rangle_2 \to (|0\rangle_1+|1\rangle_1)|0\rangle_2/2+|1\rangle_1|1\rangle_2/\sqrt{2}$. Now, if a measurement made on the computer switch gives $|0\rangle_1$ (i.e., the measurement collapses the state to $|0\rangle_1|0\rangle_2$), then one can conclude that the answer to the computation is *not* '0' since the computer switch cannot be in state $|0\rangle_1$ when the answer is '0'. A Mitchison-Jozsa type history analysis shows that there is only one history contributing to this outcome, and in this history, the computer *does not* run (see Table-IVa for this history).

This is only part of the story, however. The definition of CFC should not give different answers as to whether or not the computation is counterfactual depending on what is inside of the computer, as long as at the end there is no distinguishing information between the running and non-running histories when the answer is '0'. Therefore, when we perform the analysis including the inner workings of the computer, we should see that the outcome is still counterfactual. Consider a simple quantum computer, with two possible answers, which makes use of interference to evaluate the answer (as is the case for any kind of quantum computer). Assume that the initial state is $|0\rangle_2$. As the first step, the input/output register is prepared in a superposition of all possible states by applying a Hadamard gate, that is $|0\rangle_2 \to (|0\rangle_2 + |1\rangle_2)/\sqrt{2}$. As the second step, an 'oracle' puts a $\pi$-phase shift on the state $|1\rangle_2$ if the answer to the computation is '1', and the 'oracle' leaves the state alone if the answer to the computation is '0'. As the last step, a Hadamard gate is applied to the qubit. The final state is $|0\rangle_2$ if the answer is '0', and $|1\rangle_2$ if the answer is '1'. This quantum computer behaves exactly as the two-answer computer that we used in the preceding paragraph.

Now, we would like to use the Mitchison-Jozsa type history analysis to determine whether the computation is counterfactual or not, taking into account the inner workings of the computer. We ask the question: Does the input/output register of the computer ever take on the value $|1\rangle_2$ in the outcome in which the final measurement on the switch is $|1\rangle_1$. This is another way of asking whether the computer has run in this particular outcome, since the input/output register taking on the value $|1\rangle_2$ can only happen while the computer is running (however, the input/output register being $|0\rangle_2$ *does not* indicate that the computer has not run). Using the Mitchison-Jozsa type history analysis, this time we arrive at Table-IVb for the particular measurement outcome we are interested in. In Table-IVb each history $h$ is a list of measurement outcomes ('real' and 'hypothetical', terms introduced by Mitchison and Jozsa [1]; we exactly follow their procedure in determining whether a particular event has happened or not in a 'real' measurement outcome). The real measurement is the measurement outcome of the switch qubit at the end, and the hypothetical measurement is a fictitious labeling measurement (not a real quantum measurement) made on the input-output register in the middle of the computer run-time. In each history list $h$, the label $n$ is used if the hypothetical measurement indicates that the input-output register is $|1\rangle_2$, and the label $u$ is used if the hypothetical measurement indicates that the input-output register is $|0\rangle_2$. Label $n$ stands for 'on', and label $u$ stands for 'uncertain', since the computer being 'off' cannot be determined from the input-output qubit being $|0\rangle_2$.

| TABLE-IV a | | |
|---|---|---|
| | $h$ | $v_h$ |
| 1: | $f\ 0_1$ | $\frac{1}{2}|0\rangle_1|0\rangle_2$ |

| TABLE-IV b | | |
|---|---|---|
| | $h$ | $v_h$ |
| 1: | $n\ 0_1$ | $-\frac{1}{4}|0\rangle_1|0\rangle_2$ |
| 2: | $u\ 0_1$ | $\frac{3}{4}|0\rangle_1|0\rangle_2$ |

Table-IVb shows that there are two histories containing the real measurement outcome $0_1$, namely $u\ 0_1$ and $n\ 0_1$. Since the latter contains an $n$, Mitchison and Jozsa's first condition for CFC is not satisfied, and there is a non-vanishing amplitude for the input/output register to be found in state $|1\rangle_2$ during the protocol (which can only happen if the computer was running) when the measurement outcome $0_1$ is obtained. Therefore the computation is *not* counterfactual.

We see that, depending on how we analyze the problem, we can arrive at contradictory answers as to whether or not the computation is counterfactual. However, we believe there is general agreement, up to now, that this original simple CFC protocol *is* indeed counterfactual. Therefore, we are again led to conclude that individual histories connecting the beginning and the end of a CFC protocol *cannot* be used to reliably answer whether or not the computation is counterfactual.

change the main point of the argument, namely, that only the dark output port of the interferometer leads to the screen. All the nearby paths going through the interferometer will destructively cancel each other in the outcome we are considering.

[5] We admit that one *could* say that a particular event has taken place prior to a measurement outcome if there is *any* quantum trajectory which involves the event (regardless of whether there are other trajectories which may cancel out with this trajectory). The problem with such a definition, however, is that it is not *useful*. Since, according to strict Feynman path formulation, the photon has histories along *every* space-time trajectory from the 'Source' to the point 'S' on the screen, it could take *any* trajectory, e.g., into an adjacent room and back. With this definition of determining whether or not an event has taken place, it seems, to us, that there is no possibility that *any* event *ever* does *not* take place.

[6] P. G. Kwiat, J. R. Mtchell, P. D. D. Schwindt, and A. G. White, *J. Mod. Opt.* **47**, 257 (2000).

[7] The rotations are defined as follows:

$|0\rangle_1 \xrightarrow{R'} \cos\theta' |0\rangle_1 + \sin\theta' |1\rangle_1$ and $|1\rangle_1 \xrightarrow{R'} -\sin\theta' |0\rangle_1 + \cos\theta' |1\rangle_1$, where $\theta' = \frac{\pi}{2N'}$

$|0\rangle_2 \xrightarrow{R} \cos\theta |0\rangle_2 + \sin\theta |1\rangle_2$ and $|1\rangle_2 \xrightarrow{R} -\sin\theta |0\rangle_2 + \cos\theta |1\rangle_2$, where $\theta = \frac{\pi}{2N}$.

Here $N'$ is the number of total routines, while $N$ is the total number of subroutines in a routine.

[8] This definition (interpretation) has its own problems, as we describe in more detail in the second part of this paper: Reply to the response of Dr. Mitchison and Dr. Jozsa. However, we still argue on physical grounds that the chained-Zeno protocol is completely counterfactual.

[9] To be precise, the area where the probability is exactly zero is a set of measure zero. It was precisely for this subtlety that we initially used interferometers.

[10] Private communication.